\documentclass{emulateapj}

\shorttitle{Gemini imaging of QSO host galaxies}
\shortauthors{Croom et al.}

\def \bj {b_{\rm J}}
\def \mb {M_{\rm b_{\rm J}}}
\def \re {R_{\rm e}}
\def \lsq {L^*_{\rm QSO}}
\def \lsg {L^*_{\rm gal}}
\def \ebv {E(B-V)}
\def \kmsmpc {{\rm ~km~s}^{-1}~{\rm Mpc}^{-1}}

\begin{document}


\title{Gemini imaging of QSO host galaxies at $z\sim2$}


\author{Scott M. Croom}
\affil{Anglo-Australian Observatory, PO Box 296, Epping, NSW, 1710, Australia}
\email{scroom@aaoepp.aao.gov.au}

\author{David Schade}
\affil{Dominion Astrophysical Observatory, 5071 West Saanich Road,
Victoria, V8X 4M6, Canada}

\author{Brian. J. Boyle}
\affil{Anglo-Australian Observatory, PO Box 296, Epping, NSW, 1710, Australia}

\author{Tom Shanks}
\affil{Department of Physics, University of Durham, South Road, Durham, DH1 3LE, UK}

\author{Lance Miller}
\affil{Department of Physics, Oxford University, 1 Keble Road, Oxford,
OX1 3RH, UK}

\and

\author{Robert J. Smith}
\affil{Astrophysics Research Institute, Liverpool John Moores
        University, Twelve Quays House, Egerton Wharf, Birkenhead,
        CH41 1LD, UK}



\begin{abstract}
We present results of a Gemini adaptive optics (AO) imaging program to
investigate the host galaxies of typical QSOs at $z\sim2$.  Our aim
is to study the host galaxies of typical, $\lsq$ QSOs at the epoch of
peak QSO and star formation activity.  The large database of faint
QSOs provided by the 2dF QSO Redshift Survey allows us to select a
sample of QSOs at $z=1.75-2.5$ which have nearby ($<12$ arcsecond
separation) bright stars suitable for use as AO guide
stars.  We have observed a sample of 9 QSOs.  The images of these
sources have AO corrected full-width at half-maximum of between 0.11
and 0.25 arcseconds.  We use multiple observations of point spread
function (PSF) calibration star pairs in order to quantify any
uncertainty in the PSF.  We then
factored these uncertainties into our modelling of the QSO  plus host
galaxy.  In only one case did we convincingly detect a host (2QZ
J133311.4+001949, at $z=1.93$).  This host galaxy has $K=18.5\pm0.2$
mag with a half-light radius, $\re=0.55\pm0.1''$ , equivalent to
$\sim3\lsg$ assuming a simple passively evolving model.  From detailed
simulations of our host galaxy modelling process, we find that for
four of our targets we should be sensitive to host galaxies that
are equivalent to $\sim2\lsg$ (passively evolved).  Our non-detections
therefore place tight constraints on the properties of $\lsq$ QSO host
galaxies, which can be no brighter (after allowing for passive
evolution) than the host galaxies of $\lsq$ AGN at low redshift,
although the QSOs themselves are a factor of $\sim50$ brighter.  This
implies that either the fueling efficiency is much greater at high
redshift, or that more massive black holes are active at high redshift.

\end{abstract}




\section{Introduction}

In recent years it has become evident that active galactic nuclei
(AGN) could play a crucial role in the formation of most galaxies.
Although only a small fraction of galaxies at any one time contain an
AGN, it is possible that all sufficiently massive galaxies passed
through an active phase at some point in their history.  Answering the
question of what triggers this activity is a vital step in gaining a
full understanding of galaxy formation.

Recent observations that most nearby luminous galaxies contain a
massive black hole (or at least a 'massive dark object') certainly
support the hypothesis that these galaxies all passed through an AGN
phase \citep[e.g.,][]{kor95}.  In particular \citet{mag98} suggest
that all dynamically hot systems (either elliptical galaxies or spiral
bulges) will contain a central massive dark object.  The measured
masses of these dark objects are found to be correlated to the
luminosity (or mass, assuming a uniform mass-to-light ratio) of the
spheroidal component, with more massive spheroids containing more
massive dark objects.  Detailed analysis has shown that black hole
mass appears to correlate most tightly with the velocity dispersion of
the spheroid \citep{geb00a,fer00}.  This relation also appears to hold
in galaxies with current nuclear activity \citep{geb00b,fer01}.

Further circumstantial evidence of a connection between AGN and
galaxy formation is that the strong increase in the space density of
luminous QSOs with increasing redshift \citep{2qz1} closely matches
the increase in the the global star formation rate \citep{bt98}.  Both
of these appear to peak at $z\sim2-3$.

\begin{table*}
\begin{center}
\caption{Parameters of observed QSOs.}\label{tab:qsodetails}
\scriptsize{
\begin{tabular}{ccccccccc}
\tableline
\tableline
QSO name &   R.A.  &  Dec.   & $\bj$ & $z$ & $\mb$\tablenotemark{a} & $\ebv$\tablenotemark{b} & GS sep.\tablenotemark{c} & GS $r$\tablenotemark{c}\\ 
     & (J2000) & (J2000) & (mag) && (mag) & &($''$) & (mag)\\
\tableline
2QZ J103204.7$-$001120 & 10 32 04.75 & $-$00 11 20.2 & 19.25 & 1.8761 & $-$26.27 & 0.064 &  9.72 & 14.7\\ 
2QZ J105113.8$-$012331 & 10 51 13.86 & $-$01 23 31.9 & 19.88 & 2.4103 & $-$26.12 & 0.044 & 11.48 & 14.5\\ 
2QZ J111859.6$-$001737 & 11 18 59.63 & $-$00 17 37.9 & 20.60 & 1.7494 & $-$24.67 & 0.049 & 11.00 & 15.3\\ 
2QZ J112839.8$-$015929 & 11 28 39.84 & $-$01 59 29.0 & 19.21 & 1.7462 & $-$26.03 & 0.041 & 10.98 & 13.8\\ 
2QZ J133311.4$+$001949 & 13 33 11.42 & $+$00 19 49.6 & 20.27 & 1.9325 & $-$25.16 & 0.023 & 11.73 & 13.8\\ 
2QZ J134441.0$-$004951 & 13 44 41.09 & $-$00 49 51.2 & 18.89 & 2.2605 & $-$26.87 & 0.024 & 11.98 & 12.8\\ 
2QZ J140854.0$-$023626 & 14 08 54.04 & $-$02 36 26.4 & 20.69 & 1.8029 & $-$24.69 & 0.058 & 11.07 & 14.2\\ 
2QZ J144115.5$-$005726 & 14 41 15.51 & $-$00 57 26.3 & 19.38 & 2.2578 & $-$26.47 & 0.047 & 10.88 & 15.8\\ 
2QZ J145049.9$+$000143\tablenotemark{d} & 14 50 49.92 & $+$00 01 43.9 & 19.37 & 1.9679 & $-$26.20 & 0.046 & 11.66 & 14.4\\ 
\tableline
\end{tabular}}
\tablenotetext{a}{Assuming $\Omega_0=0.3$, $\Lambda_0=0.7$ and $H_0=70\kmsmpc$.}
\tablenotetext{b}{Galactic $\ebv$ taken from Schlegel et al (1998).}
\tablenotetext{c}{Guide star magnitude and separation from QSO.}
\tablenotetext{d}{FIRST radio detection of $6.51\pm0.14$ mJy.}
\end{center}
\end{table*}

Parameters such as the shape, size and luminosity of AGN host galaxies
can help to determine how the activity occurs, and also shed light on
the process of galaxy formation.  The first observations of AGN host
galaxies focused on low redshift sources, limited as they were by the
moderate resolution of most ground based facilities.  It has been know
for some time that radio-loud (RL) sources are exclusively found in
elliptical galaxies while low-luminosity radio-quiet (RQ) Seyfert
galaxies were thought to prefer spiral galaxies.  However, more recently
\citet{tay96} found that almost half of the RQ AGN in
their sample had early-type hosts.

The high resolution available with the Hubble Space Telescope (HST)
has provided significant advances in the measurement of AGN host
galaxies.  At low redshift ($z<0.15$) \citet{sch00} have imaged the
hosts of 76 AGN selected from the Einstein Extended Medium Sensitivity
Survey.  These should be free of any selection bias with respect to
host properties, and lie in a luminosity range bracketing the
extrapolated break in the AGN luminosity function,
$M^*_{B(AB)}\simeq-22$.  In this sample of RQ AGN 55\% had hosts
which were dominated by a spheroidal component.  Apart from this bias
towards earlier morphological types, in all other respects the host
galaxies are identical to normal galaxies.  For example, they follow
the same size-luminosity relations for disks and spheroids as normal
galaxies.  For brighter low-redshift AGN the hosts are invariably
elliptical galaxies \citep{mcl99}.

At higher redshift, the most systematic and extensive work so far has
been by \citet{kuk01}, who have observed RQ and
RL AGN with $-24\leq M_{\rm V}\leq -25$ at $z=1$ and $z=2$ in
the rest-frame V-band using NICMOS on the HST.  At $z=1$ they find
hosts which are consistent with the passive evolution of elliptical
galaxies.  At higher redshift ($z\sim2$) it appears that the
RQ AGN have hosts which are are somewhat fainter than a
passively evolving model, while the RL AGN hosts are still
consistent with passively evolved elliptical galaxies.  However,
although they made detections of hosts at $z=2$, Kukula et
al. struggled to fit both size and luminosity for their $z=2$ RQ
sample, as they where faint and compact (similar in size to the NICMOS
PSF in the $H$-band).

Our aim in this paper is to make a detailed investigation of the host
galaxy properties of AGN at $z\sim2$, near the peak in both AGN and
cosmic star-formation activity.  Our approach has been to select AGN
for study that are near to $\lsq$ in luminosity.  This
therefore allows us to examine the hosts of {\it typical} AGN at high
redshift.  In order to obtain sufficient signal-to-noise ($S/N$) and
spatial resolution we have used the infrared adaptive optics (AO)
system, Hokupa'a, at Gemini North to carry out an imaging survey of
high  redshift AGN hosts.  At $z\sim2$, $L^*$ AGN typically have
apparent  magnitudes of $B\sim19-20$.  This is too faint to be used as an
AO guide star.  Therefore we have used the large database of
the 2dF QSO Redshift Survey (2QZ; Croom et al. 2001;
Croom et al. 2004) to select AGN which are nearby ($<12''$ separation)
bright Galactic stars which can act as AO guide stars.

In Section \ref{sec:sample_obs} we discuss our sample, observations
and data reduction.  In Section \ref{sec:image_analysis} we present
analysis of our final AO corrected images, and in Section
\ref{sec:fitting} fit multi-component models to the data, as well as
carrying out simulations to determine our expected sensitivity
limits.  We discuss our conclusions in Section \ref{sec:discuss}

\begin{figure}
\plotone{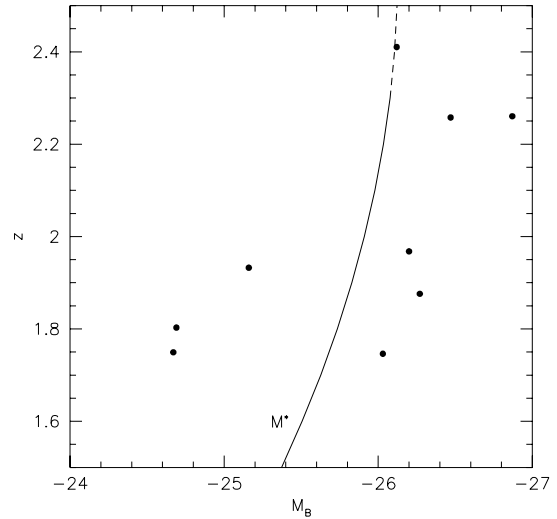}
\caption{The distribution of our sample in the redshift-absolute
magnitude plane.  The solid line shows the location of the break in the
QSO luminosity function, $\mb^*$, as a function of redshift taken from
the luminosity function model fits of Boyle et al. (2000) and Croom et
al. (2002).  Beyond $z=2.3$ the position of $\mb^*$ is extrapolated
(dotted line).\label{fig_zmag}}
\end{figure}

\begin{table*}
\begin{center}
\caption{The parameters of our PSF stars.}\label{tab:psfdetails}
\begin{tabular}{cccccc}
\tableline
\tableline
PSF star name & R.A. & Dec. & $r$\tablenotemark{a} & GS sep.\tablenotemark{b} & GS $r$\tablenotemark{c} \\ 
& (J2000) & (J2000) & (mag) &($''$) & (mag)\\
\tableline
J103204.7$-$001120-PSF & 10 53 51.76 & $-$00 13 51.3 & 15.4 &  9.97 & 15.2\\ 
J105113.8$-$012331-PSF & 10 51 27.40 & $-$02 57 20.2 & 16.2 & 12.56 & 14.9\\ 
J111859.6$-$001737-PSF & 11 20 58.16 & $-$02 03 33.4 & 15.7 & 11.76 & 15.6\\ 
J112839.8$-$015929-PSF & 11 56 22.70 & $+$01 39 56.2 & 16.3 & 11.35 & 13.4\\ 
J133311.4$+$001949-PSF & 12 33 37.08 & $-$00 42 07.4 & 16.2 & 12.22 & 14.5\\ 
J134441.0$-$004951-PSF & 14 14 05.66 & $+$01 11 27.0 & 13.3 & 12.69 & 16.3\\ 
J140854.0$-$023626-PSF & 13 59 05.93 & $-$01 15 50.0 & 16.5 & 11.74 & 14.6\\ 
J144115.5$-$005726-PSF & 14 35 51.85 & $-$03 16 35.7 & 16.1 & 10.60 & 16.0\\ 
J145049.9$+$000143-PSF & 14 28 47.90 & $-$02 14 42.6 & 14.4 & 12.64 & 15.8\\ 
\tableline
\end{tabular}
\tablenotetext{a}{Photographic $r$ magnitude of PSF star.} 
\tablenotetext{a}{Separation in arcseconds of PSF star and guide star.} 
\tablenotetext{c}{Photographic $r$ magnitude of guide star.} 
\end{center}
\end{table*}

\section{Observations and data}\label{sec:sample_obs}

\subsection{Sample selection}

The 2QZ is a color selected QSO sample based on photographic data
from the UK Schmidt Telescope.  QSOs are selected by their blue
stellar appearance from $u$, $\bj$ and $r$ photographic plates and
films.  The survey comprises 30 UKST fields arranged in two
$75\degr\times5\degr$ declination strips centered on
$\delta=-30\degr$ and $\delta=0\degr$.  The $\delta=-30\degr$
strip extends from $\alpha=21^h40$ to $\alpha=3^h15$ in the South
Galactic Cap and the equatorial strip from $\alpha=9^h50$ to
$\alpha=14^h50$ in the North Galactic Cap.  The range in apparent
magnitude is $18.25<\bj<20.85$, and QSOs are selected up to $z\sim3$.
Details of the catalogue along with the public release are given in
Croom et al. (2004).  Details of the photometric candidate selection
are given in Smith et al (2004).

The 2QZ catalogue contains over 23000 QSOs.  At the time of
our observations it contained 13000 QSOs, of which approximately 6000
were in the $\delta=0\degr$ strip and therefore easily visible from Gemini
North.  To obtain good AO correction using the Hokupa'a system on
Gemini we were required to have guide stars which were brighter than
$R=16$ and had a maximum separation from the QSO of $12''$.  Also, we
wish to examine QSO host galaxies at high redshift, at or near the
peak epoch of QSO activity and star formation.  We therefore also
limited ourselves to QSOs with $z=1.75-2.5$.  Searching our
photometric catalogue for QSOs in this redshift range with a nearby
bright star resulted in a sample of $\sim20$ potential targets.  Of
these, nine where actually observed with Gemini.  The details of the
target QSOs are listed in Table \ref{tab:qsodetails}.  Their absolute
magnitudes are calculated assuming a cosmological model with
$\Omega_0=0.3$, $\Lambda_0=0.7$ and $H_0=70$~km~s$^{-1}$~Mpc$^{-1}$
(used throughout this paper) and using the K-corrections of
\citet{cv90}.  We have also corrected for Galactic extinction based on
the work of \citet{sfd98}.  The distribution of redshifts and absolute
magnitudes, $\mb$, are displayed in Fig. \ref{fig_zmag}.  Also shown
(solid line) is the location of the break in the QSO luminosity
function (LF), $\mb^*$, for the best fit evolutionary model to the 2QZ
LF \citep{2qz1,2qz9}.  Our sample was specifically chosen to span the
region around $\mb^*$ to sample typical QSOs.  Our data set was
selected purely on it observed frame optical properties, with no
reference to emission properties at X-ray or radio wavelengths.
However, we have cross-correlated our sources with the FIRST
\citep{first97} 21cm radio survey and only one of our targets is
detected, J145049.9$+$000143, which has a flux of $6.51\pm0.14$~mJy.
All other targets must have radio fluxes of less than $\sim1$~mJy at
21cm.  


The second part of the source selection was to choose suitable pairs
of stars to use in our modelling of the PSF.  In particular, the
detailed form of the AO corrected PSF could depend an a number of
parameters, most notably the brightness of the AO guide star, the
separation of the guide star from the QSO and the intrinsic,
pre-correction, seeing.  A first order estimate of the PSF can be
obtained simply by using the image of the AO guide star used to
correct the QSO images, however this does not account for variations
due to the QSOs being off axis (although it does correct perfectly for
both the brightness of the AO guide star and the intrinsic seeing).
To model more exactly the PSF, we have selected a set of bright star
pairs.  One of each pair is matched to the properties of a specific AO
guide star used for the QSO observations, in both separation from the
target source and brightness.  At bright magnitudes, $R\sim16$ and
brighter, the photometric calibration of the UKST plate is poor, due to
the small number of calibrating stars, and the strong saturate of the
photographic plates, being good to only $\sim0.3-0.4$ mags.  We
therefore also chose the matching PSF stars to be from the same UKST
plates as the QSOs, so that their photometric calibration would be
uniform.  The details of the PSF stars are listed in Table
\ref{tab:psfdetails}.

\subsection{Gemini observations}

Observations were obtained at the Gemini
North telescope during the nights of April 20--April 25 2001
using the University of Hawaii (UH)
Hokupa'a 35-element adaptive optics system and the QUIRC camera
\citep{gra98}.  The QUIRC camera contains a $1024\times1024$ HAWAII
HgCdTe array, with 19.98 milli-arcsecond pixels, delivering a 20 arcsec
field of view.  Expert assistance was provided by the UH team together
with Kathy Roth and John Hamilton.  Our observations where primarily
carried out in the $K'$ band with a small number of objects also
being observed in the $H$ band.  Dome flats and dark frames were
taken at the start of each night and at the end of some
nights. Standard star observations from the UK Infrared Telescope
faint standards (FS) list were obtained each night.  A key aim in
making our observations was to obtain high quality estimates of the
PSF.  Exposures were taken in a 9-point grid dither pattern with
spacings of approximately $4''$, making sure that the AO guide star
was always on the detector.  We also selected exposure times in order
to avoid saturating QSOs, AO guide stars and PSF stars (in a single
case the PSF was saturated; observations of J144115.5$-$005726 on
April 24).  Observations of PSF stars were interleaved between science
observations to allow us to track the varying atmospheric conditions.

Table \ref{tab:obsdetails} lists the details of our observations.  The
number of frames, the total exposure time on source are listed
together with the FWHM measured.  A number of objects were repeated on
several nights in order to obtain longer exposures or improved image
quality.

\begin{table*} 
\begin{center}
\caption{Details of our Gemini observations.}
\label{tab:obsdetails}
\begin{tabular}{lccrrcl}
\tableline
\tableline
Target name             &    Date  & Filter& No. of & Time & FWHM  & Notes\\
                        &          &       & frames & (s)  & ($''$)& \\
\tableline
J112839.8$-$015929*    &  April 20 & $K'$  &  54   & 6357 &  0.16 & 2nd PSF        \\ 
J134441.0$-$004951*    &  April 20 & $K'$  & 144   & 4320 &  0.15 & 2nd PSF        \\ 
J145049.9$+$000143*    &  April 20 & $K'$  &  20   & 1200 &  0.16 &                \\ 
J112839.8$-$015929*    &  April 22 & $K'$  &   8   &  960 &  0.25 & 2nd PSF        \\ 
J112839.8$-$015929     &  April 22 & $H$   &  44   & 6240 &  0.33 & 2nd PSF        \\ 
J134441.0$-$004951     &  April 22 & $H$   & 141   & 4830 &  0.24 & 2nd PSF        \\ 
J105113.8$-$012331     &  April 23 & $K'$  &  45   & 2700 &  0.25 & QSO on edge    \\ 
J140854.0$-$023626*    &  April 23 & $K'$  & 157   & 4710 &  0.15 & Binary GS      \\ 
J145049.9$+$000143*    &  April 23 & $K'$  &  63   & 3780 &  0.19 &                \\ 
J105113.8$-$012331*    &  April 24 & $K'$  &  54   & 2430 &  0.11 &                \\ 
J111859.6$-$001737*    &  April 24 & $K'$  &  63   & 5535 &  0.18 &                \\ 
J133311.4$+$001949*    &  April 24 & $K'$  &  81   & 5400 &  0.15 &                \\ 
J144115.5$-$005726     &  April 24 & $K'$  &  78   & 5890 &  0.22 & Saturated PSF  \\ 
J103204.7$-$001120*    &  April 25 & $K'$  &  81   & 4860 &  0.14 & 2nd PSF        \\ 
J111859.6$-$001737     &  April 25 & $H$   &  27   & 2700 &  0.15 & poor flat fielding\\ 
J134441.0$-$004951*    &  April 25 & $K'$  &  90   & 2700 &  0.11 & 2nd PSF (edge) \\ 
J144115.5$-$005726*    &  April 25 & $K'$  &  35   & 2100 &  0.12 &                \\ 
\tableline
J134441.0$-$004951-PSF &  April 20 & $K'$  &  27   &  135 &  0.12 &                \\ 
J134441.0$-$004951-PSF &  April 22 & $H$   &  27   &  135 &  0.23 & 2 PSFs         \\ 
J105113.8$-$012331-PSF &  April 23 & $K'$  &  36   &  855 &  0.24 &                \\ 
J140854.0$-$023626-PSF &  April 23 & $K'$  &  27   &  540 &  0.16 &                \\ 
J145049.9$+$000143-PSF &  April 23 & $K'$  &  27   &  540 &  0.20 &                \\ 
J105113.8$-$012331-PSF &  April 24 & $K'$  &   9   &  180 &  0.11 &                \\ 
J111859.6$-$001737-PSF &  April 24 & $K'$  &   9   &  540 &  0.16 &                \\ 
J133311.4$+$001949-PSF &  April 24 & $K'$  &  12   &  720 &  0.15 &                \\ 
J144115.5$-$005726-PSF &  April 24 & $K'$  &  11   &  660 &  0.23 &                \\ 
%
\tableline
\end{tabular}
\tablecomments{Objects with a $*$ after their name are included in our
analysis.  Those not included are generally rejected due to poor image
quality (e.g. saturated PSF).}
\end{center}
\end{table*}



There were a wide range of  observing
conditions or instrument performance which varied on a short
time-scale, often showing dramatic changes in the resolution within a
single set of integrations. Thus the resulting data set has a range of
sensitivities with respect to detection of host galaxies and this
requires detailed modelling of the sensitivities.


\subsection{Data Reduction}

Reduction of the images from Hokupa'a/Quirc was done using parts of
the IRAF GEMINI.GEMTOOL and GEMINI.QUIRC packages. Dome flat fields
with a range of exposure times were taken and the task
GEMINI.QUIRC.QFLAT task was used to  make flat field images and bad
pixel masks. 

Exposure times for science images ranged from 30 to 240 seconds so
that many images (16 to 200) were taken and combined to form the final
images.  The main constraint on exposure time was the requirement that
the AO guide star did not saturate because that star serves as the
first order estimate of the PSF.  Each cycle of the 9-point dither
pattern took about 30 minutes and a separate sky was derived for each
cycle from the science frames themselves using the QSKY task ensuring
that objects were masked beyond the radius at which the faint wings
were detectable, typically $\sim2''$.  The combination of this large
masking radius, and the $\sim4''$ jitter pattern spacing means that on
scales $>2''$ residual host galaxy flux might be suppressed if it were
present.  However, reasonable physical models of host galaxies, do not
have significant flux at this radius, and even if they did, they would
then be easily detectable at smaller angular scales.  The sky from
each dither cycle was subtracted from each member image of that
cycle. The sky subtraction and flat-fielding was accomplished using
the QREDUCE task.

The stacking of images was done with scripts developed specifically
for this purpose.  The images were inspected visually, and shifts
determined by centroiding on the AO guide star.  We note that using
the QSO image to centroid on would have resulted in large random
errors due to the relatively low S/N of the QSO image in a single
frame.  A small number ($\sim 1\%$) of very poor quality images
(usually due to bad guiding) were rejected at this stage.  After sky
subtraction, individual images were shifted and then averaged, scaling
by exposure time (as there was no significant variation in sky level
which would have necessitated a variance weighted combination).

We analysed the variations in flux due to airmass within a large (30
pixel, $0.6''$) aperture, but no trend was found.  We therefore simply
scaled our images by exposure time.  Weights in the combination
process were also proportional to exposure time.

We found that best results were obtained if a separate sky was
constructed for each distinct exposure time within a set of
integrations rather than combining different exposure times into a
single sky.

\subsection{Standard stars}

Standard stars were measured with a series of apertures from 0.06 to 8
arcseconds and the growth curves examined.  Zero-points were defined
with reference to the Mauna Kea Observatory system in $K$ and $H$,
using apertures of $2''$ which was confirmed as reasonable 
from the growth curves.  No significant
extinction term was found in the $K'$ band.  In the $H$ band,
observations were not made over sufficient range in airmass to derive
an extinction term.  Therefore no extinction term was applied to
either the $K'$ or $H$ band photometry. 

\section{Image analysis}\label{sec:image_analysis}

In total 17 observations were made of 9 science targets.  Of these,
three (J112839.8$-$015929, J134441.0$-$004951 and J111859.6$-$001737)
were observed in the $H$-band, as well as $K'$.  the aim of these
$H$-band observations was to determine the colors of host galaxies,
if detected.  As host galaxies were not detected in these sources and
the FWHM of the $H$-band PSFs were generally worse than in the
$K'$-band, we restrict our discussion below to concern  the $K'$-band
imaging.  We therefore consider the 14 $K'$-band images of 9 science
targets.  Five of our targets had repeats observations, these were
taken to obtain images in better seeing conditions
(J134441.0$-$004951, J105113.8$-$012331, J144115.5$-$005726) or with
longer exposure times (J145049.9$+$000143).  Of these 14 observations
a further 2 were rejected:  The observation of J105113.8$-$012331 on
the 23rd April has the QSO very close to the edge of the detector in
one of the dither positions; The April 24th observation of
J144115.5$-$005726 has a saturated PSF star.  Removing these
from our sample we have a set of 12 observations of 9 QSOs.  The
number of individual exposures that were combined to  produce a single
image ranged from 8 to 157 (median 36) with total exposure times that
ranged from 960 seconds to 6357 seconds with a median of 4320
seconds. The total integration times for individual objects (where
more than 1 stack was produced for an object) ranged from 4700 to 7990
seconds in $K'$ (median 5400 seconds) and ranged from 2700 to 6240
seconds for the three objects that were also observed in $H$.  Image
quality in the aligned and stacked science images  ranged from 0.11 to
0.25 arcseconds full-width at half maximum (FWHM) in the $K'$-band,
with a median of 0.15 arcseconds.  A list of the observational details
appears in Table \ref{tab:obsdetails}.

Also listed in Table \ref{tab:obsdetails} are the observations of PSF
stars made in order to calibrate the PSFs in our science images.
Eight separate observations were made of PSF stars in the $K'$-band,
and one in the $H$-band.  These observations were made in order for us
to better characterize the expected PSF of our QSO targets, and hence
to make more accurate model fits to determine the host galaxy
contribution to the source flux.

\subsection{Point-spread function analysis}\label{sec:psf}

Before analysing the QSOs, we require a detailed characterization of
the PSF.  A central part of the problem of detecting and measuring host
galaxies around these AGN is understanding the PSF. The challenges of
modelling the PSF are severe with AO systems. The PSF is expected to
change as a function of distance from the AO guide star.  Increasing
distance from the guide star degrades the resolution, and some
elongation is expected, with the major axis being directed towards the
location of the guide star.  Variations will also be expected,
depending on the brightness of the AO guide star, and the uncorrected
seeing.  Our observations were designed to enable us to quantify these
effects.

Preliminary investigations showed that Moffat profiles provided a
significantly better fit to the PSFs than did Gaussians.  Therefore,
in our analysis below, we will use Moffat profiles of the form
\begin{equation}
I=I_{\rm c}(1+r^2/\alpha^2)^{-\beta},
\end{equation}
where $I$ is the intensity at a given pixel, a distance $r$ from the
center of the source, $I_{\rm c}$ is the central intensity of the
source, and $\alpha$ and $\beta$ are parameters which define the shape
of the profile.  $\beta$ is typically $\simeq2$ for our images.  We
will also discuss the FWHM of the images, which for a Moffat profile
is defined as
\begin{equation}
FWHM=2\alpha(2^{1/\beta}-1)^{1/2}.
\end{equation}  
We also investigated whether there was any evidence of a core-halo
structure in our PSFs.  Even when stacking multiple PSFs we found no
evidence of an extended halo around the compact core of the PSF.  This
is most likely attributable to the relatively low strehl ratios
($<0.1$) in our observations.

In all cases the images of our science targets included the AO guide
star on the chip. This provides a first order measurement of the PSF
obtained in the same conditions as the science observation. In several
cases there exist secondary PSF stars on the science frames in
addition to the guide star. These secondary PSFs provide an estimate
of the spatial variation of the PSF with distance from the guide
star. In the case that no secondary PSF stars were available in the
AO-corrected field a star-star pair was observed to quantify the
spatial variation of the PSF. One of the stars served as the AO guide
star and primary PSF and the second served as the secondary PSF.  Our
8 PSF star observations had a median separation of 12.6 arcseconds
(range 11.1 to 13.3 arcseconds), and the 5 star-star pairs found on
the science images had separations from 6.0 to 14.5 arcseconds with a
median of 10.6 arcseconds. For comparison, the separations from the AO
guide star to the science targets ranged from 9.9 to 13.0 arcseconds
with a median of 10.7 arcseconds. The "delivered" resolution of the
aligned and stacked PSF images ranged from 0.11 to 0.24 arcseconds
with a median value of 0.15 arcseconds. This is similar to the
distribution of science image delivered resolutions.

The fundamental measurements that provide a basic characterization of
the PSF are the FWHM of the profile, the ellipticity, and the position
angle (PA). These were derived by fitting a Moffat profile to the
two-dimensional image using the IRAF IMEXAMINE task. The "direct" FWHM
was adopted.

\begin{figure}
\epsscale{0.85}
\plotone{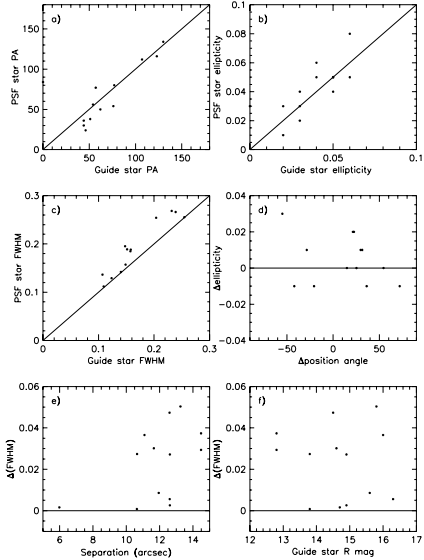}
\caption{The distribution of PSF properties derived from PSF stars
and secondary PSF stars found in QSO fields.  The first three plots
show a comparison of AO guide star and PSF star properties for a)
position angle, b) ellipticity, c) FWHM.  The next three plots show
the difference between AO guide star properties and PSF star
properties for d) ellipticity difference vs. position angle
difference, e) FWHM difference vs. distance between the AO guide star
and the PSF star, f) FWHM vs. guide star $R$-band
magnitude.\label{fig:psf}}
\end{figure}

Fig. \ref{fig:psf} is a comparison of the parameters of all of the
star-star pairs in our observations both from those science frames that
had secondary PSFs and from the star-star pairs that we observed.  We
first directly compare the PA, ellipticity and FWHM of the
guide stars and the PSF stars in Figs. \ref{fig:psf}a, b and c.  We
see that the AO guide stars and PSF stars generally have the same
PAs and ellipticity, with no evidence of systematic changes
in orientation or ellipticity seen for the PSF stars.  As in all cases
the sources are close to circular, the error on the PA is
large.  In Fig. \ref{fig:psf}c we show the FWHM of the PSF stars
vs. that of the AO guide stars.  We see that the PSF stars generally
have a slightly wider FWHM than the AO guide stars, which is as
expected given that the PSF stars are off axis.

We next show the quantity $\Delta$PA in Fig. \ref{fig:psf}d.  This is
the absolute value of the difference between
the PA of the model fit to the secondary PSF star and
angle of a line from the secondary PSF to the AO guide star.  We plot
this against the difference in ellipticity between the AO guide star
and the PSF star.  If the PSF star was elongated with its major axis
pointing toward the guide star then these angles should cluster around
zero, that is, the major axis should tend to point toward the guide
star.  This is not the case.  On the same figure the quantity
$\Delta$ellipticity is the difference in ellipticity between the AO
guide star and the PSF star. If the PSF star was elongated
then this should tend to be a positive quantity. This is not the
case.  We therefore fail to detect any elongation of the PSF stars, in
the direction of the AO guide star, or any other direction.

Figs. \ref{fig:psf}e and f show the difference between the FWHM of
the PSF star and the AO guide star
($\Delta$FWHM=(FWHM(PSF)-FWHM(GS))).  A positive value
indicates that the resolution of the PSF star is degraded
relative to the guide star. The expected degradation is detected and
ranges from zero (no degradation) to 0.05 arcseconds or a relative
loss of resolution of roughly 30\%. It might be expected that the most
significant difference between guide star FWHM and PSF FWHM would
occur with the faintest guide stars and the largest separation between
the PSF stars and AO guide stars.  We see no significant evidence that
this is the case.  The closest PSF star (at $\sim6''$ separation) does
have a $\Delta$FWHM which is close to zero, however, at larger
separations other PSF stars with $\Delta$FWHM close to zero can be
found.  The AO guide star - QSO separations are $9.9-13.0''$, and in
this range there is no observable trend of $\Delta$FWHM with
separation.  Over the range of AO guide star magnitudes used for our
targets there is also no evidence of any correlation between guide
star magnitude and $\Delta$FWHM.

In summary, we see that there is typically a degradation of
$\sim0.03''$ (full range of $0-0.05''$) at the separations used in our
observations.  However the amount of degradation does not seem
to depend on any observable parameters.  This dispersion in
$\Delta$FWHM will be accounted for in our analysis below.

\begin{figure*}
\epsscale{.80}
\plotone{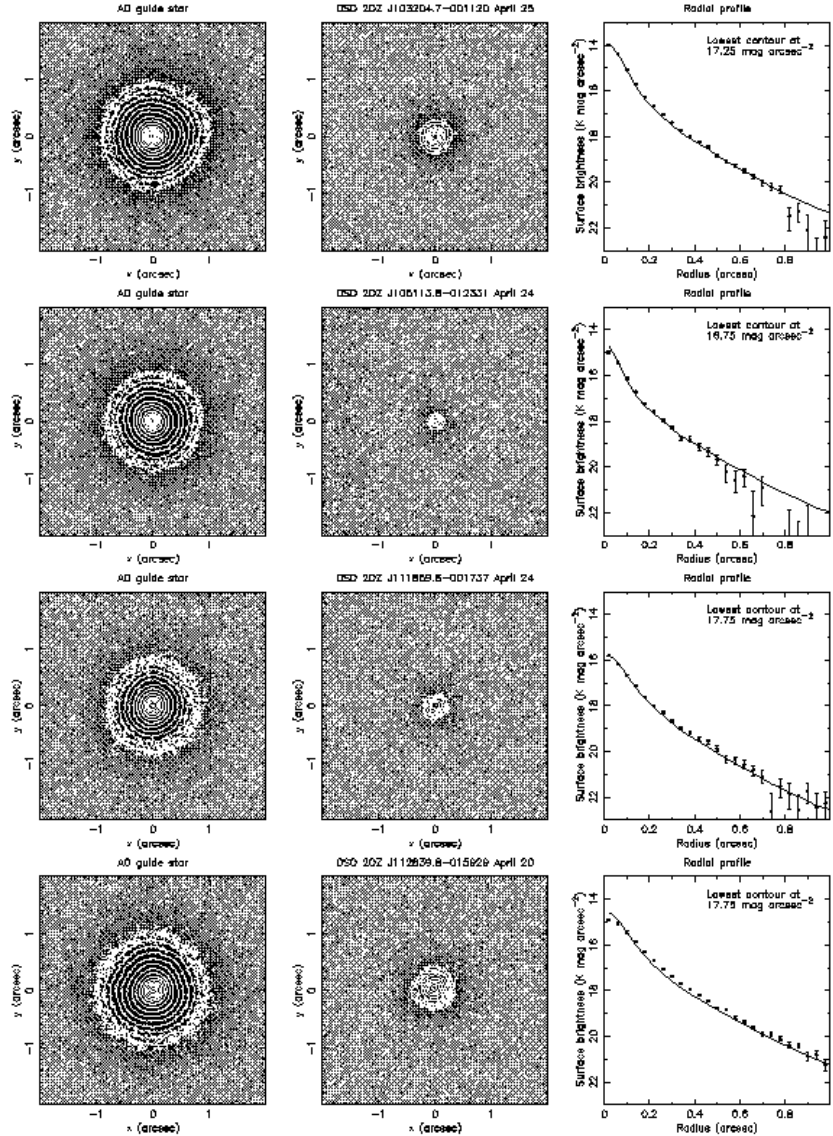}
\caption{AO guide star (left) and QSO (center) images from our
$K'$-band observations, together with the measured radial profiles
(right).  The contours are in 0.5 mag arcsec$^{-2}$ intervals, and the
lowest contour is the first above 5 times the RMS sky noise.  The
surface brightness of the lowest contour is shown in the radial
profile plot (right) for each observation.  The radial
profile of the AO guide star (right; solid line) is normalized to the
QSO (points), using the total flux within 0.2 arcsec.  Errors in the QSO
radial profile are derived from the scatter in each radial bin, and do
not include errors in the PSF.\label{fig:qsoim}}
\end{figure*}

\addtocounter{figure}{-1}

\begin{figure*}
\epsscale{.80}
\plotone{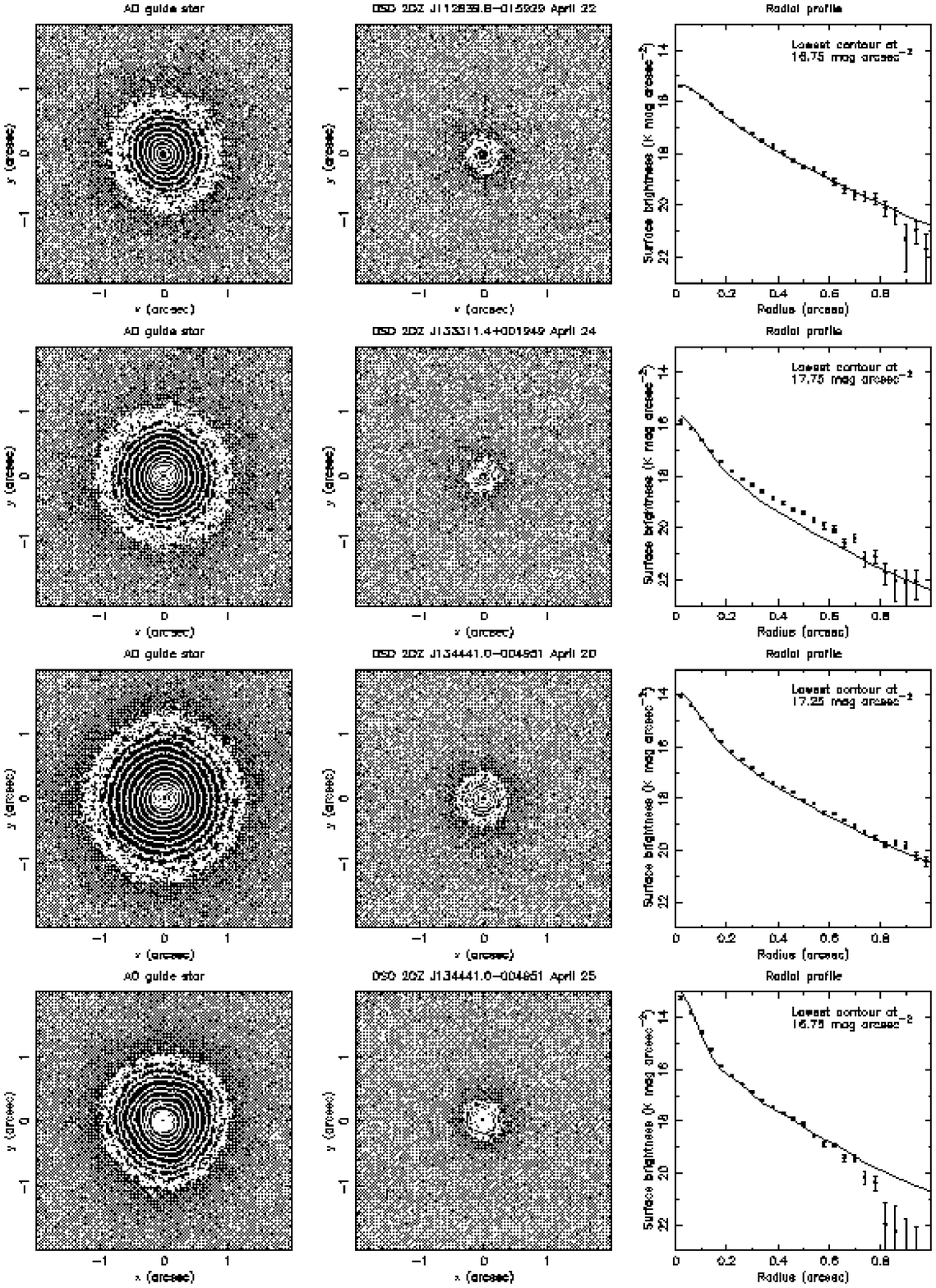}
\caption{continued.}
\end{figure*}

\addtocounter{figure}{-1}

\begin{figure*}
\epsscale{.80}
\plotone{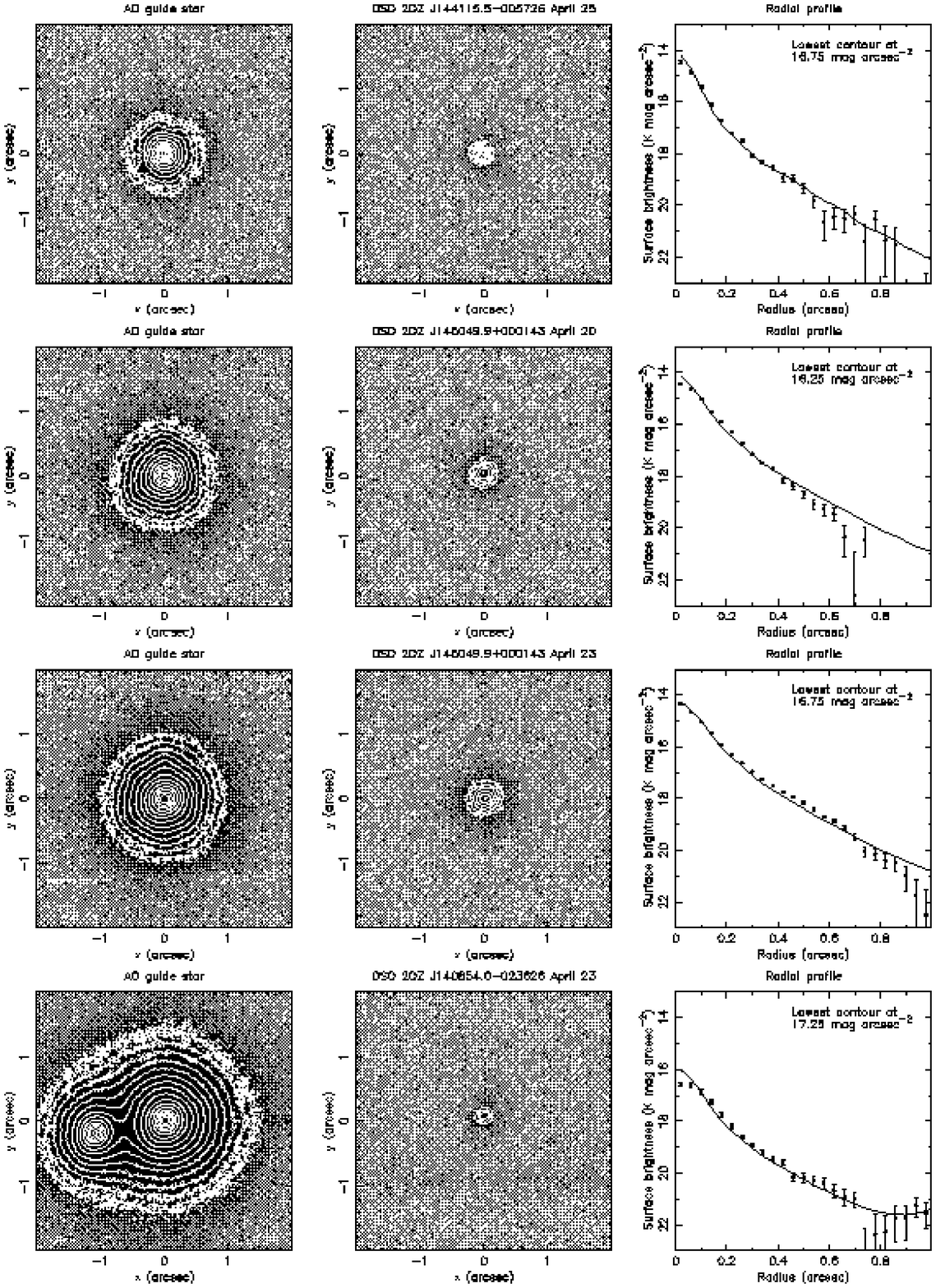}
\caption{continued.}
\end{figure*}

\subsection{QSO images}

Given the above estimates of the PSF variations, we now attempt to
model the QSO images, in order to derive the fundamental parameters of the host
galaxies.  The images of the 12 AO guide star/QSO pairs are shown in
Fig. \ref{fig:qsoim}.  Both images are displayed on the same scale,
and the contours are at 0.5 mag arcsec$^{-2}$ intervals.  The level of
the lowest contour is set to be the first contour above 5 times the
RMS sky noise.  Also plotted are the radial profiles for both the AO
guide stars and QSOs.  The profiles are determined relative to the
image centroid, and are averaged in 2 pixel bins.  The errors are the
standard errors derived from the dispersion in each bin, and do not
take into account errors in the PSF.  The AO guide star profile (solid
line) has been normalized to the QSO, based on the total flux within a
0.2 arcsec radius aperture.  Although we have not broadened the AO
guide star PSF, we see that in many cases it matches closely the PSF
of the QSO.  This demonstrates that in most cases the AO guide star
gives a good estimate of the PSF.  By a straight forward comparison of
the profiles, we can see that few of the QSOs show evidence for an
extended component.  The QSO 2QZ J133311.4$+$001949 is the best
example of a source with an extended component, with the QSO radial
profile being significantly broader that the AO guide star PSF.  In
our model fitting below we will assess the significance of this
extended component and attempt to determine its parameters.

\section{Two-dimensional profile fitting}\label{sec:fitting}

If the form of the host galaxy luminosity profile is known reasonably
well then an effective way to detect the presence of a host galaxy is
to do a simultaneous fit of the nuclear point source and the host
galaxy luminosity profile including the convolution with the PSF. This
procedure can also yield reliable limits on the brightness and other
properties of possible sources in the case of no detection. This
procedure requires assumptions about the possible range of
morphologies of sources.  Most nearby massive galaxies have luminosity
profiles that are reasonably well-described by some combination of an
exponential disk and a more compact component, often modelled using
the deVaucouleurs $r^{1/4}$ law form. \citet{sch00} extended this
approach by adding a point-source component and the results suggest
that the characteristics of the underlying host galaxy can be
extracted reliably given a sufficient combination of signal-to-noise
ratio and spatial resolution. This approach is adopted for the present
study.

\subsection{Fitting procedure}

The details of the fitting procedure are similar to those described in
\citet{sch00}. For each image a fit was made with the three components
described above. Given the relationship between spheroids and massive
central objects \citep{geb00a,fer00} we are particularly interested in
the detection of bulge-like components.  Models were integrated over each
pixel and convolved with the adopted PSF. Our observational procedure
provides a first-order estimate of the PSF (based on the AO guide
star) and our separate PSF star observations provide an estimate of
the change in the PSF from the guide star position to the position of
the QSO. As shown in Fig. \ref{fig:psf} we observe scatter in the
quantity $\Delta (FWHM)$ (the change in the full-width at half maximum
of a point source between the guide star position and the science
target position)  rather than systematic variations in FWHM and the
other properties (ellipticity and major axis position angle) that
could be used to make precise predictions of the characteristics of
the PSF at the science target position. In the absence of a predictive
trend, our best estimate of the PSF at the science target positions is
the guide star PSF broadened by $\sim0.03$  arcseconds.  With this
assumption we expect $\sim 80\%$ of the observations to have FWHM that
lie within 0.02 arcseconds of that predicted resolution. 

In practice, we deal with the PSF uncertainty by fitting each science
image with three PSFs.  These cover a range of FWHM values from that
of the observed AO guide star to 0.04--0.05 arcseconds greater than
the AO guide star.  This range of $\Delta (FWHM)$ is approximate, as
we chose actual PSFs from other observations to use as the broadened
PSFs, rather than trying to artificially broaden the AO guide star
PSF.  In a small number of cases PSFs of the required FWHM were not
avaiable, and so for these objects, PSFs were constructed based on the
observed AO guide star using the IRAF MAGNIFY task.  The management of
the PSFs was done using DAOPHOT \citep{stetson} which models the PSF
by a combination of an analytical function and a look-up table of
residuals. This PSF can then be centered at any position with respect
to the pixel grid.

\subsection{Simulations of host galaxy detection sensitivity}

In order to understand the effectiveness of the fitting procedure an
extensive set of simulations were performed where images of galaxies
with known size, surface brightness, and morphology were created and
then subjected to the fitting procedure and the outcome evaluated. In
particular, we estimated the probability of detecting  elliptical host
galaxies of luminosities $\lsg$, $2\lsg$, and $3\lsg$. At redshift
$z=2$ a $2\lsg$ elliptical galaxy subject to passive evolution
(assuming a single burst of star formation at $z=5$) corresponds to an
observed $K'=19$ mag (being $\sim1.3$ mag more luminous in the rest
frame $I$-band than at $z=0$).  Luminosities of $\lsg$ and $3\lsg$
correspond to $K'$-band apparent magnitudes of 19.8 and 18.6
respectively.

The relation between size and luminosity is taken from Schade,
Barrientos, \& Lopez-Cruz (1997) and corresponds to half-light radius
($R_e$) of 0.37, 0.65, and 0.90 arcseconds for $\lsg$, $2\lsg$, and
$3\lsg$ respectively. For each source, the observed $K'$
magnitude was taken to be the sum of a nuclear point source and a host
galaxy with pure bulge morphology (no disk component) with luminosity
of $\lsg$, $2\lsg$, or $3\lsg$ and this assumption determined the
division of light between the bulge and nuclear point-source
components. The sources in our sample with the brightest apparent
magnitudes are completely dominated by the point source component and
any host galaxy would thus be very difficult to detect unless it were
extremely luminous. Pure bulge components for the host galaxy
represent the most difficult case for detection by the fitting process
because disk components would be expected, in general, to be more
extended at a given luminosity and thus easier to separate from the
point source. Thus if the host galaxies had disk morphologies then the
probability of detecting them would be higher at a given luminosity
than the estimates for elliptical hosts.

Simulations were prepared on a source-by-source basis using the actual
exposure times, sky brightnesses (which varied significantly), and
noise characteristics for the images of each object (we also included
repeat observations apart from the April 22 observation of
J112839.8--015929 which has a short integration time and poor
seeing).  In each case, a model of the final deep QSO image was added
to a suitable sky frame (rather than simulating each frame of our
dithered observations).
The relation between the resolution of the AO guide star (our first
order PSF) and the actual resolution at the position of the QSO was
estimated from those frames where two or more stars were available on
the frame (see Section \ref{sec:psf} above).  Centering errors were
introduced into the simulations by assigning fractional-pixel
positions that varied randomly so that each galaxy was centered
differently with respect to the pixel grid. Then the centroiding
algorithm was applied to the simulated images to determine the center
for the purposes of the fitting procedure. This is a relevant source
of error because the PSF and the models are both discrete before
convolution and a discrete fast Fourier transform is used to
accomplish the convolution.

\begin{figure}
\epsscale{1}
\plotone{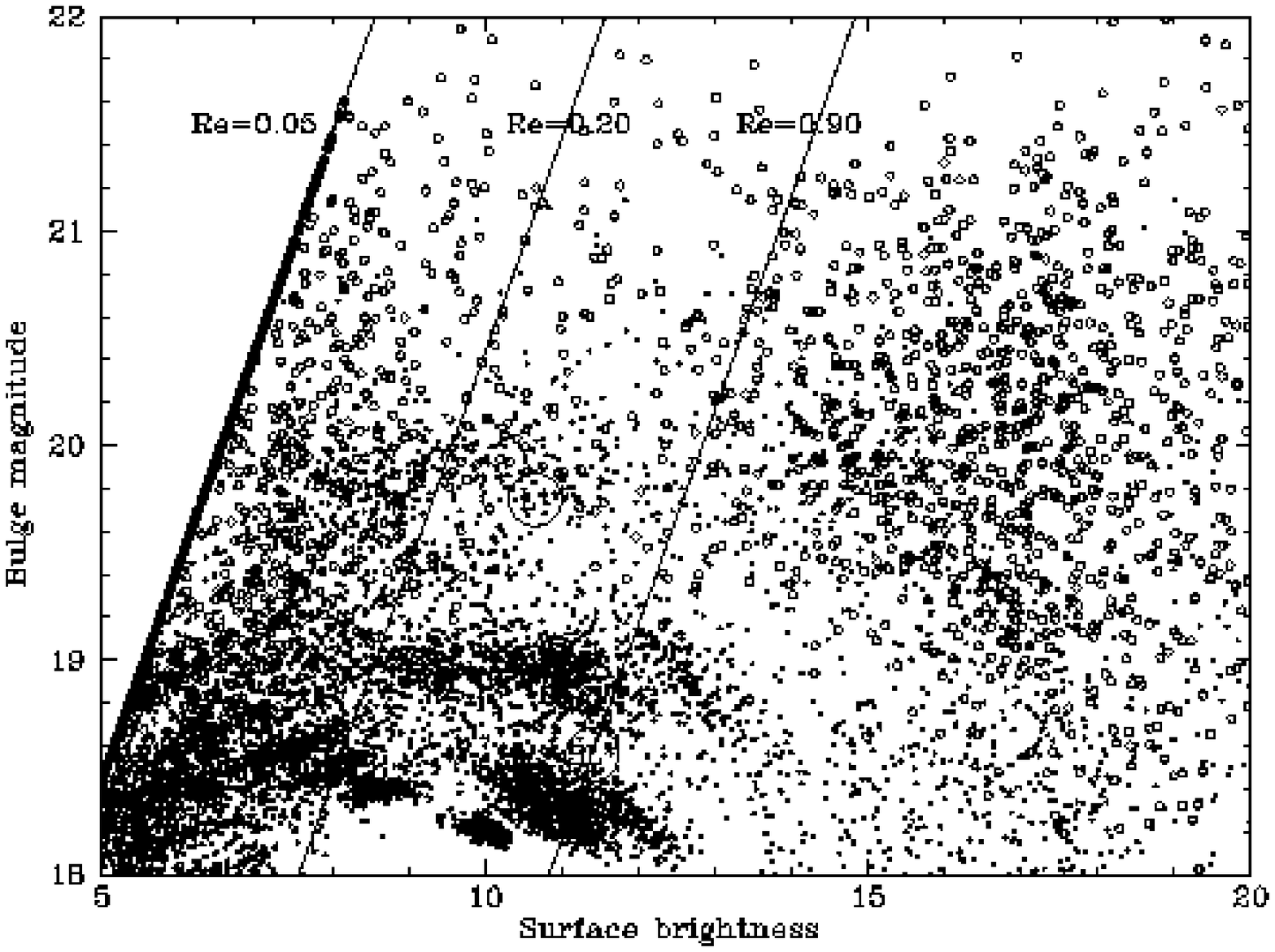} 
\caption{The results of applying the fitting process to simulated
images.  We show the total bulge magnitude vs. surface brightness (in
magnitudes per arcsec$^2$) derived from the fits to the simulated
images  The open circles show the results of the
fitting procedure for 5500 simulated images where no galaxy was
present, that is, the simulation was a point source only. These
point-source results include the full range of errors in the PSF. The
filled circles are for simulations where galaxies were present, also
with full range of PSF errors.  
The clumps of
points near (surface brightness, magnitude) equal to (10.0, 19) and
(10.5, 18.5) are detections of simulated galaxies with luminosities
$2\lsg$ and $3\lsg$ respectively.\label{fig:MuMag}}
\end{figure}

The uncertainty in the PSF is a major source of error in detecting
galaxies and estimating their sizes and surface brightness. Each set
of 100 simulations was repeated with 5 different PSFs which varied in
FWHM around the true value from 0.02 arcseconds smaller to 0.02
arcseconds larger than the value that was used to create the simulated
galaxy image. Therefore, each of our observations of acceptable
quality was simulated 1500 times (100 simulations for each of 3 model
galaxy luminosities and each of 5 PSFs).  Each observation was also
simulated 500 times  (100 simulations for each of 5 PSFs) with only a
point source. Fig.  \ref{fig:MuMag} shows the results of these
simulations. The point source simulations (open circles) are shown for
the full range of PSF simulations as are the galaxies (filled  circles).
The clumps of
points near (surface brightness, magnitude) equal to (10.0, 19) and
(10.5, 18.5) are detections of simulated galaxies with luminosities
$2\lsg$ and $3\lsg$ respectively. Few galaxies at $\lsg$ were
detected. Two thirds (67.6\%) of the simulated $\lsg$ galaxies were
fitted with half-light radii at the fitting limit of 0.05 arcseconds.
From this figure it is clear that our observations have little
sensitivity to elliptical galaxies fainter than K=19.5 and reasonable
structural parameters.  Note that much of the plane in Fig.
\ref{fig:MuMag} represents unphysical sets of parameters for galaxies
in the local Universe.

For the purposes of estimating sensitivity a successful detection will
be defined as a fitting result giving a galaxy brighter than $K'=20$ and
with a half-light radius larger than 0.2 arcseconds.  Further, we will
require that the surface brightness in $K'$ is less than 14 magnitudes
per square arcsecond. With this definition we find globally that 1.2
per cent of the simulations of pure point sources are detected as
galaxies so that the rate of spurious detections is small even
including the effect of PSF errors.

Table \ref{tab:sims} gives the FHWM and total $K'$ magnitdue for each
observation together with the detection probabilities assuming an
elliptical host galaxy of the given luminosity. Probabilities are
shown for the case where the PSF is known exactly and for a case where
there are uncertainties in the PSF. For the latter case the average is
weighted with the case of a perfect PSF given a weight of unity and
errors of 0.01 and 0.02 in the FWHM of the PSF given weights of 0.5
and 0.25 respectively. These weights are a crude way to include the
fact that the larger errors are less likely to occur than the smaller
errors. We have insufficient information to do better than this
estimate.

The detection probabilities range from 0 to 1 and only a fraction
of our observations have significant sensitivity to the galaxies that
we have modelled.  Our observations have very little sensitivity to
galaxies less luminous that $\lsg$.  This is largely due to their
small size ($R_e=0.37''$, compared to $R_e=0.65''$ for a $2\lsg$
galaxy).  The sum of the probabilities at each
luminosity can be taken as the expected number of galaxies that would
have been detected at a given luminosity if all QSOs resided in
host galaxies at that luminosity. Using the weighted probability we
would have expected 0.4 galaxies of luminosity $\lsg$ to be detected,
2.6 galaxies at $2\lsg$, or 4.0 galaxies of $3\lsg$ under the
assumptions stated. The expected number of detected galaxies is
slightly higher if the PSF uncertainties are negligible (0.2, 3.3, and
4.6 galaxies respectively).  In some cases, when the detection
probability is low, the probability is greater when allowing for PSF
errors.  This is due to spurious detections.  We also note that the
four objects for which we have reasonable probabilities of detecting
host galaxies are the faintest four sources in our sample.  Contrast
clearly plays an important role in being able to distinguish a faint
bulge component, as the point source becomes brighter a bulge
component of a given brightness becomes harder to detect.    

\begin{table*}
\begin{center}
\caption{Detection Probabilities.}\label{tab:sims}
\begin{tabular}{cccccccc}
\tableline
\tableline
Name  & Night & FWHM   & Exp.     & Mag      & $P(\lsg)$ & $P(2\lsg)$ & $P(3\lsg)$  \\
      &       & ($''$) & Time (s) & ($K'$)   &           &            &             \\ 
\tableline			      				   				  
J112839.8$-$015929 & April 20 &   0.16 & 6357  & 17.004 & 0.00 (0.01) & 0.01 (0.06) & 0.01 (0.10) \\
J134441.0$-$004951 & April 20 &   0.15 & 4320  & 16.348 & 0.00 (0.00) & 0.00 (0.00) & 0.00 (0.00) \\
J145049.9$+$000143 & April 20 &   0.16 & 1200  & 17.070 & 0.01 (0.04) & 0.01 (0.06) & 0.09 (0.20) \\
J140854.0$-$023626 & April 23 &   0.15 & 4710  & 18.052 & 0.01 (0.06) & 0.81 (0.64) & 1.00 (0.90) \\
J105113.8$-$012331 & April 24 &   0.11 & 2430  & 17.887 & 0.15 (0.06) & 0.80 (0.32) & 1.00 (0.42) \\
J111859.6$-$001737 & April 24 &   0.18 & 5535  & 18.343 & 0.00 (0.12) & 0.85 (0.79) & 1.00 (1.00) \\
J133311.4$+$001949 & April 24 &   0.15 & 5400  & 18.077 & 0.00 (0.05) & 0.84 (0.64) & 1.00 (0.91) \\
J144115.5$-$005726 & April 24 &   0.22 & 5890  & 17.395 & 0.00 (0.00) & 0.00 (0.00) & 0.00 (0.00) \\
J103204.7$-$001120 & April 25 &   0.14 & 4860  & 16.919 & 0.00 (0.00) & 0.00 (0.01) & 0.00 (0.04) \\
J134441.0$-$004951 & April 25 &   0.11 & 2700  & 16.348 & 0.00 (0.00) & 0.00 (0.00) & 0.02 (0.00) \\
J144115.5$-$005726 & April 25 &   0.12 & 2100  & 17.395 & 0.00 (0.03) & 0.02 (0.10) & 0.46 (0.42) \\
\tableline                                   						  
\end{tabular}
\tablecomments{For each object the probabilities of detecting a
  $\lsg$, $2\lsg$ and $3\lsg$ bulge are given assuming perfect
  knowledge of the PSF, with the probabilities allowing for PSF errors
  in parentheses.}
\end{center}
\end{table*}

\subsection{Fitting results}

We have calculated the probabilities for detection of galaxies with a
very specific set of assumptions about their size, luminosity, and
morphology.  The simulations demonstrated that spurious detections
occur some small fraction of the time largely because  of the
uncertainty in the PSF. We will adopt the same set of criteria for
accepting a host galaxy detection as legitimate as we chose from the
simulations. We require a detection of a host galaxy brighter than
$K'=20$ and with a half-light radius larger than 0.2 arcseconds and with
a $K'$-band surface brightness less than 14 mag~arcsec$^{-2}$.

As in the simulations, many of the fits resulted in the statistical
detection of a galaxy component in addition to the point source but
most  of these "detections" are galaxies with sizes at the limit of
0.05 arcseconds. In these cases the galaxy component is simply
assuming the role of fitting small residuals between the adopted PSF
and the actual PSF (which is not precisely known). The fitting
procedure is not sensitive to host galaxies with sizes that are so
small that they approach point sources after accounting for the
convolution with the PSF. 

\subsubsection{Detections}

Fits to the actual observations resulted in a number of apparent
detections.  These are discussed below:

\subsubsection{J112839.8$-$015929}

The observation of J112839.8$-$015929 on the 22nd April showed a
detection of a bulge with a reasonable set of galaxy parameters. Both
the size and surface brightness fall within our acceptable range for a
range of PSFs, with a best fit $R_e\simeq1.2$ arcseconds.  However, the
observation was of poor quality ($0.25''$ FWHM and short exposure
time, see Fig. \ref{fig:qsoim}).  There are two other images of this
object and both were of  superior quality (one in the $H$-band). Fits
to those images resulted in no detection.  This detection is thus
rejected as spurious but it is worrying that we needed additional data
to reject it.

\begin{figure}
\plotone{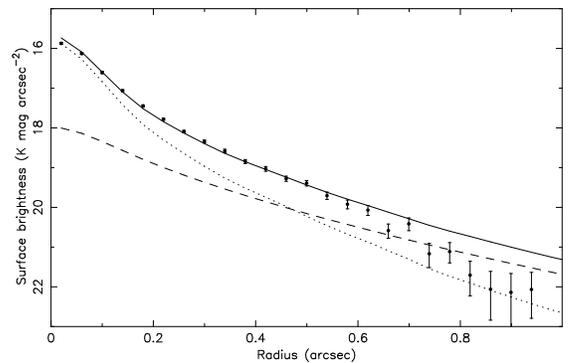} 
\caption{The radial profile of 2QZ J133311.4$+$001949 (points)
compared to a model fit (solid line) comprising a point source (dotted
line) and bulge (dashed line) component.  The bulge component has been
convolved with the PSF as measured from the AO guide star in this
field.\label{fig:rad}}
\end{figure}

\subsubsection{J133311.4$+$001949}

The best detection that we have is on the observation of
J133311.4$+$001949 on the 24th April.  The data quality is good
(FWHM=$0.15''$ and exposure time 5400 seconds). The detection is
fairly robust against changes in the fitting PSF. The best fit
parameters for the host galaxy are $K'=18.5\pm0.2$, half-light radius 
$R_e=0.55\pm0.1$ arcseconds and central surface brightness in $K'$ of
$12.3\pm0.3$. The errors are estimates from the range of 
best-fit parameters after varying the PSF. Visual examination of the
fit yields no reason to reject it.  The fitted parameters would
indicate that this is a galaxy of luminosity roughly 3L* with a
slightly higher surface brightness and smaller radius than
expected.  The radial surface brightness distribution for
J133311.4$+$001949 is shown in Fig. \ref{fig:rad}.  The errors shown
are purely statistical, based on the scatter within each annulus.
Overplotted is the best fit model (solid line) which comprises a point
source (dotted line) and an $r^{1/4}$ law bulge convolved with the PSF
derived from the AO guide star (dashed line).  A clear excess over the
point source profile is seen which is well described by an $r^{1/4}$
law at $<0.7''$.  

\begin{figure}
\plotone{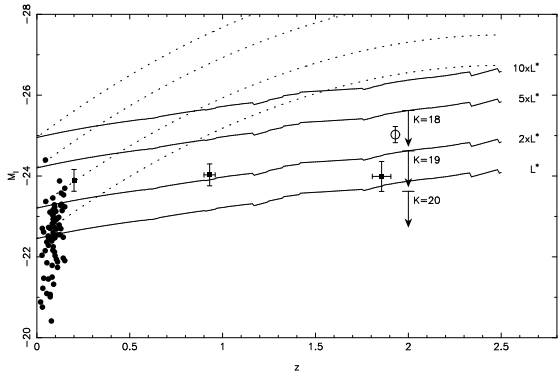} 
\caption{The variation of QSO host galaxy luminosity with redshift.
We show the rest-frame absolute $I$-band magnitudes, $M_I$ as a
function of redshift.  The expected passive evolution of a single
instantaneous burst stellar population formed at $z=5$ is shown by the
solid lines  for various luminosities, relative to $L^*$.  As a
comparison we also show the host galaxy luminosities if they evolved
similarly to QSO luminosity (dotted lines).  At low
redshift, $z<0.2$, are the data from \citet{sch00}, and at $z\simeq2$
the arrows denote upper limits corresponding to $K=18$, 19 and 20 (top
to bottom), transformed to the rest-frame $I$.  We plot our single
detection of a host galaxy at $z=1.93$ (open circle) and compare this
to the results of Kukula et al. (2001) for radio quiet QSOs (filled
squares).\label{fig_evol}}
\end{figure}

\section{Discussion}\label{sec:discuss}

Given the upper limits to potential host galaxy magnitudes derived in
the simulations above.  We now consider what these imply for the
evolution of QSO host galaxies.  Our observations were made in the
$K$-band, which is close to the rest-frame $I$-band.  This then allows
direct comparison with the low redshift observations of
\citet{sch00}.  We wish to determine whether our observations are
consistent with host galaxies which evolve more or less strongly that
may be predicted by simple passive evolution.  To determine this we
make comparisons to the simple single instantaneous burst models of
Bruzual \& Charlot (1995).  In Fig. \ref{fig_evol} we show the
evolution of a single instantaneous burst at $z=5$ in the rest-frame
$I$-band (solid lines) and compare this to the measured luminosities
of QSO host galaxies (similar to Fig. 7 of Kukula et al. 2001).  The
evolutionary curves are relative to the local value of $\lsg$ found
from the 2dF Galaxy Redshift Survey \citep{n00}.  The evolution is
shown for galaxies of current day luminosity $\lsg$, $2\lsg$, $5\lsg$
and $10\lsg$.  At low redshift we plot the host galaxy magnitudes
found by  Schade et al. from their low redshift X-ray selected sample
(filled circles).  Although there is large scatter (which is partly
due to a weak correlation with luminosity), it is clear that these
host galaxies have luminosites which are at most a few times $\lsg$.
We also plot the average absolute magnitudes of radio quiet QSO host
galaxies measured by Kukula et al (2001) (filled squares) transformed
from the $V$ band to the $I$ band using the $V-I$ color of the
passively evolving model at each redshift.  These show
no evidence of any evolution with redshift.  Our host galaxy detection
at $z=1.93$ is shown (open circle), and is seen to lie at an absolute
magnitude equivalent to $\sim3\lsg$ (passively evolved).  We also mark
the location of upper limits corresponding to $K =18$, 19 and 20 at
$z\simeq2$ (arrows).  The above simulations showed that our host
galaxies must typically have $K$-band magnitudes fainter than
$K\sim18.5-19.0$,  therefore they are consistent with being only a few
times the passively evolved $\lsg$.  As a comparison we also show the
evolution found in the AGN point source (dotted line), as measured by
the evolution of the QSO luminosity function from the 2QZ
\citep{2qz1}.  If the host galaxies were to evolve by this amount,
they would be easily detectable in our Gemini observations.

Thus, our Gemini observations show that the host galaxies of high
redshift QSOs cannot be significantly more luminous than would be
predicted from passive evolution of their stellar population.
Although we have used a simple model (a single instantaneous burst) to
describe host galaxy passive evolution, it is worth noting that any
more complex model (e.g. with on-going star-formation) is likely to
show much stronger evolution so long as star formation declines
towards low redshift.  This agrees with the HST observations of Kukula
et al. (2001) which also show little evolution in the host galaxy
properties of radio-quiet QSOs.  We do note however, that Kukula et al
were not able to constrain both the size and brightness of their
detected host galaxies in their radio quiet sample.   

We can also view this from the perspective of the relationship 
between bulge mass and black hole mass derived for AGN at $z=0$ 
(Magorrian et al.\ 1998, Gebhardt et al.\ 2000a).  Casting this 
in terms of bulge and AGN luminosity we obtain (see McLeod, Reike \&
Storrie-Lombardi 1999; Schade et al. 2000):


\begin{eqnarray}
M_{I_{\rm AGN}}= & M_{I_{\rm Bulge}}-4.8-2.5\log\epsilon-2.5\log\left(
\frac{\Upsilon_I}{10M/L}\right)\nonumber \\ 
 & -2.5\log\left(\frac{f}{0.002}\right)+2.5\log\left(\frac{BC}{10}\right) 
\end{eqnarray}

where $\epsilon$ is the AGN luminosity expressed in terms of the Eddington
luminosity, $\Upsilon_I$ is the mass-to-light ratio in the $I$ band,
BC is the bolometic correction from $I$ to total luminosity for the AGN 
and $f$ in the fraction of spheroid mass in the black hole.

Adopting values consistent with low redshift AGN; $\epsilon=0.01$, 
$BC=10$ and $\Upsilon_I=10$, the relation reduces to 

\begin{equation}
M_{I_{\rm AGN}}=M_{I_{\rm Bulge}}+0.2-2.5\log\left(\frac{f}{0.002}
\right)
\end{equation}

In this analysis, we obtain an upper limit of $M_{I_{\rm Bulge}}>-24.3$
for a typical $L^*$ QSO: $M_{I_{\rm AGN}}\sim -27.0$, implying that 

\begin{equation}
\log\left(\frac{f}{0.002}\right)>1
\end{equation}

i.e. that the fraction of black hole mass to bulge mass is at least an
order of magnitude higher at $z\sim2$ than at $z\sim0$.  Equally, 
an increase in the efficiency of the black hole to $\epsilon\sim0.1$  could also
account for the observed upper limit on the bulge luminosity whilst
maintaining $f=0.002$.  However, in either case, the faintness of 
the host galaxy at $z\sim 2$ in relation to observations at low redshift
suggests that the conditions in AGN at $z\sim 2$ are fundamentally different
to those studied at lower redshift; requiring at least an order of
magnitude change in the Eddington luminosity ratio or in the mass of the
black hole in relation to the bulge mass of the galaxy.

Moreover, a recent analysis of QSO spectra from the 2QZ shows that the
black hole mass to AGN luminosity relation does not appear to evolve
strongly with redshift over the range $0<z<1.5$ (Corbett et al.\
2003).  This observation implies little if any change in the value of
$\epsilon$ over this redshift range, suggesting that it is 
the evolution of black hole mass (and thus the bulge-to-black hole
mass ratio), rather than fuelling rate, that is driving the observed
evolution of the QSO luminosity function.  Thus, more massive black
holes are active at high redshift.  This rather counter-intuative
picture suggests that QSOs would
have a single active phase at the formation of the spheroid, the
subsequent evolution in the QSO LF reflecting the decreasing efficiency
with which massive black holes could be formed in a spheroid of given
mass over the range $0<z<2$.  This clearly needs further
investigation, as the evolution of the black hole mass to
bulge mass correlation at high redshift is a crucial to understanding 
the formation of QSOs.  Once clear detections of host galaxies have
been made, follow up to determine black hole mass (e.g. through
reverberation mapping) should be of high priority.







\acknowledgments

Based on observations obtained at the Gemini Observatory, which is
operated by the Association of Universities for Research in Astronomy,
Inc., under a cooperative agreement with the NSF on behalf of the
Gemini partnership: the National Science Foundation (United States),
the Particle Physics and Astronomy Research Council (United Kingdom),
the National Research Council (Canada), CONICYT (Chile), the
Australian Research Council (Australia), CNPq (Brazil) and CONICET
(Argentina).  This paper is based on observations obtained with the
Adaptive  Optics System Hokupa'a/Quirc, developed and operated by the
University of Hawaii Adaptive Optics Group, with support from the
National Science Foundation.  We thank Kathy Roth, John Hamilton and
the University of Hawaii Adaptive Optics Group for their expert
assistance in carrying out our observations.  The 2QZ is
based on observations made with the Anglo-Australian Telescope and the
UK Schmidt Telescope.

\clearpage

\end{document}